\documentclass[aps,prd,twocolumn,showpacs,preprintnumbers,superscriptaddress,amsmath,amssymb]{revtex4}
\usepackage{graphicx}
\usepackage{dcolumn}
\usepackage{color}
\usepackage{epsfig} 
\graphicspath{{ps}}

\newcommand{\gev} {\ensuremath{\, {\mathrm{GeV}}     }}
\newcommand{\mev} {\ensuremath{\, {\mathrm{MeV}}     }}

\newcommand{\gevc}{\ensuremath{\, {\mathrm{GeV}/c^2} }}
\newcommand{\mevc}{\ensuremath{\, {\mathrm{MeV}/c^2} }}

\newcommand{\ecm} {\ensuremath{ E_{\mathrm{c.m.}} }}
\newcommand{\gisr}{\ensuremath{ \gamma_{\mathrm{ISR}} }}
\newcommand{\sqs} {\ensuremath{ \sqrt{s} }}
\newcommand{\RMS} {\ensuremath{ M^2_{\mathrm{recoil}} }}
\newcommand{\st}  {\ensuremath{ \mathrm{(stat.)} }}
\newcommand{\sy}  {\ensuremath{ \mathrm{(sys.)}  }}
\newcommand{\br}  {\ensuremath{ \mathcal{B} }}
\newcommand{\gi}  {\ensuremath{ \Gamma }}

\newcommand{\bee} {\ensuremath{ \br_{\mathrm{ee}} }}

\newcommand{\ee}   {\ensuremath{ e^+e^- }}
\newcommand{\dn}   {\ensuremath{ D^0 }}
\newcommand{\dpl}  {\ensuremath{ D^+ }}
\newcommand{\dstm} {\ensuremath{ D^{*-} }}
\newcommand{\dstn} {\ensuremath{ D^{*0} }}
\newcommand{\ps}   {\ensuremath{ \psi(4415) }}
\newcommand{\y}    {\ensuremath{ Y(4260) }}

\newcommand{\dnpip}   {\ensuremath{ \dn        \pi^+ }}
\newcommand{\ddstp}   {\ensuremath{ \dn  \dstm \pi^+ }}
\newcommand{\dpdstmp} {\ensuremath{ \dpl \dstm \pi^+ }}
\newcommand{\dstdstp} {\ensuremath{ D^{*+} \overline D{}^{*0} \pi^+ }}

\newcommand{\dpdstm}   {\ensuremath{ D^+ D^{*-}     }}
\newcommand{\dstpdstm} {\ensuremath{ D^{*+} D^{*-}  }}
\newcommand{\dndmpp}   {\ensuremath{ D^0 D^- \pi^+  }}

\newcommand{\dd}       {\ensuremath{ D   \overline D      }}
\newcommand{\ddst}     {\ensuremath{ D   \overline D{}^* }}
\newcommand{\dndst}    {\ensuremath{ \dn   \overline D{}^{*0}}}
\newcommand{\dstdst}   {\ensuremath{ D^* \overline D{}^* }}
\newcommand{\lala}     {\ensuremath{ \Lambda_c^+ \Lambda_c^- }}

\newcommand{\eedd}      {\ensuremath{ \ee \to \dd   }}
\newcommand{\eeddp}     {\ensuremath{ \ee \to \dn D^- \pi^+ }} 
\newcommand{\eell}      {\ensuremath{ \ee \to \lala }}
\newcommand{\eeddstp}   {\ensuremath{ \ee \to \ddstp }} 
\newcommand{\eeddstpg}  {\ensuremath{ \eeddstp \gisr }}
\newcommand{\eeddstppg} {\ensuremath{ \ee \to \ddstp \pi^0_{\mathrm{miss}} \gisr }}
\newcommand{\eeddstpgg} {\ensuremath{ \ee \to \dstn \dstm \pi^+ \gisr }}
\newcommand{\eeisoch}   {\ensuremath{ \ee \to \dpl \dstm  \pi^+ \pi^-_{\mathrm{miss}} \gisr }}
\newcommand{\misid}     {\ensuremath{ \eeddstp \pi^0  }}

\newcommand{\mdn}    {\ensuremath{ M_{\dn}      }}

\newcommand{\mdstm}  {\ensuremath{ M_{\dstm} }}

\newcommand{\mddstp} {\ensuremath{ M_{\ddstp}}}

\newcommand{\pp}      {\ensuremath{\psi(2S)   }}

\newcommand{\ppjpsi}  {\ensuremath{\pi^+\pi^- J/\psi }}
\newcommand{\pppsip}  {\ensuremath{\pi^+\pi^- \pp    }}

\newcommand{\psiddt} {\ensuremath{\ps \to D \overline D{}^{*}_2(2460) }}

\begin{document}

\title{\quad\\[0.5cm] Measurement of the \eeddstp\ cross section using
  initial-state radiation}

\affiliation{Budker Institute of Nuclear Physics, Novosibirsk}
\affiliation{Chiba University, Chiba}
\affiliation{University of Cincinnati, Cincinnati, Ohio 45221}
\affiliation{T. Ko\'{s}ciuszko Cracow University of Technology, Krakow}
\affiliation{The Graduate University for Advanced Studies, Hayama}
\affiliation{Gyeongsang National University, Chinju}
\affiliation{Hanyang University, Seoul}
\affiliation{University of Hawaii, Honolulu, Hawaii 96822}
\affiliation{High Energy Accelerator Research Organization (KEK), Tsukuba}
\affiliation{Hiroshima Institute of Technology, Hiroshima}
\affiliation{Institute of High Energy Physics, Chinese Academy of Sciences, Beijing}
\affiliation{Institute of High Energy Physics, Vienna}
\affiliation{Institute of High Energy Physics, Protvino}
\affiliation{Institute for Theoretical and Experimental Physics, Moscow}
\affiliation{J. Stefan Institute, Ljubljana}
\affiliation{Kanagawa University, Yokohama}
\affiliation{Institut f\"ur Experimentelle Kernphysik, Universit\"at Karlsruhe, Karlsruhe}
\affiliation{Korea University, Seoul}
\affiliation{Kyungpook National University, Taegu}
\affiliation{\'Ecole Polytechnique F\'ed\'erale de Lausanne (EPFL), Lausanne}
\affiliation{Faculty of Mathematics and Physics, University of Ljubljana, Ljubljana}
\affiliation{University of Maribor, Maribor}
\affiliation{Max-Planck-Institut f\"ur Physik, M\"unchen}
\affiliation{University of Melbourne, School of Physics, Victoria 3010}
\affiliation{Nagoya University, Nagoya}
\affiliation{Nara Women's University, Nara}
\affiliation{National Central University, Chung-li}
\affiliation{National United University, Miao Li}
\affiliation{Department of Physics, National Taiwan University, Taipei}
\affiliation{H. Niewodniczanski Institute of Nuclear Physics, Krakow}
\affiliation{Nippon Dental University, Niigata}
\affiliation{Niigata University, Niigata}
\affiliation{University of Nova Gorica, Nova Gorica}
\affiliation{Novosibirsk State University, Novosibirsk}
\affiliation{Osaka City University, Osaka}
\affiliation{Panjab University, Chandigarh}
\affiliation{University of Science and Technology of China, Hefei}
\affiliation{Seoul National University, Seoul}
\affiliation{Shinshu University, Nagano}
\affiliation{Sungkyunkwan University, Suwon}
\affiliation{School of Physics, University of Sydney, NSW 2006}
\affiliation{Excellence Cluster Universe, Technische Universit\"at M\"unchen, Garching}
\affiliation{Tohoku Gakuin University, Tagajo}
\affiliation{Tohoku University, Sendai}
\affiliation{Department of Physics, University of Tokyo, Tokyo}
\affiliation{Tokyo Metropolitan University, Tokyo}
\affiliation{Tokyo University of Agriculture and Technology, Tokyo}
\affiliation{IPNAS, Virginia Polytechnic Institute and State University, Blacksburg, Virginia 24061}
\affiliation{Yonsei University, Seoul}

  \author{G.~Pakhlova}\affiliation{Institute for Theoretical and Experimental Physics, Moscow} 
  \author{H.~Aihara}\affiliation{Department of Physics, University of Tokyo, Tokyo} 
  \author{K.~Arinstein}\affiliation{Budker Institute of Nuclear Physics, Novosibirsk}\affiliation{Novosibirsk State University, Novosibirsk} 
  \author{T.~Aushev}\affiliation{\'Ecole Polytechnique F\'ed\'erale de Lausanne (EPFL), Lausanne}\affiliation{Institute for Theoretical and Experimental Physics, Moscow} 
  \author{A.~M.~Bakich}\affiliation{School of Physics, University of Sydney, NSW 2006} 
  \author{V.~Balagura}\affiliation{Institute for Theoretical and Experimental Physics, Moscow} 
  \author{E.~Barberio}\affiliation{University of Melbourne, School of Physics, Victoria 3010} 
  \author{A.~Bay}\affiliation{\'Ecole Polytechnique F\'ed\'erale de Lausanne (EPFL), Lausanne} 
  \author{K.~Belous}\affiliation{Institute of High Energy Physics, Protvino} 
  \author{V.~Bhardwaj}\affiliation{Panjab University, Chandigarh} 
% ...
\collaboration{The Belle Collaboration}

\begin{abstract}
We report measurements of the exclusive cross section for \eeddstp\ as
a function of center-of-mass energy from the \ddstp\ threshold to
5.2\gev\ with initial-state radiation. No evidence is found for
$\y\to\ddstp$ decays.  The analysis is based on a data sample
collected with the Belle detector at or near a center-of-mass energy
of 10.58\gev\ with an integrated luminosity of $695\,\mathrm{fb}^{-1}$
at the KEKB asymmetric-energy \ee\ collider.
\end{abstract}


\maketitle
\setcounter{footnote}{0}

Studies of exclusive open charm production near threshold in
\ee\ annihilation provide important information on the dynamics of
charm quarks and on the properties of the $\psi$ states.  During the
past three years numerous measurements of exclusive \ee\ cross
sections for charmed hadron pairs have been reported.  Most of these
measurements were performed at $B$-factories using initial-state
radiation (ISR).  Belle presented the first results on the \ee\ cross
sections to the \dd, \dpdstm~\cite{chcon}, \dstpdstm,
\dndmpp\ (including the first observation of
\psiddt\ decays)~\cite{belle:dd,belle:dst,belle:4415} and \lala\ final
states~\cite{belle:x4630}. BaBar measured \ee\ cross sections to
\dd\ and recently to the \ddst, \dstdst\ final states~\cite{babar:dd,
  babar:dd_new}.  CLEO-c performed a scan over the energy range from
3.97 to 4.26\gev\ and measured exclusive cross sections for the \dd,
\ddst\ and \dstdst\ final states at thirteen points with high
accuracy~\cite{cleo:cs}. The measured open charm final states nearly
saturate the total cross section for charm hadron production in
\ee\ annihilation in the \sqs\ region up to $\sim 4.3\gev$.  In the
energy range above $\sim 4.3\gev$ some room for contributions to the
\ps\ state from unmeasured channels still remains. The exclusive cross
sections for charm strange meson pairs have been measured to be an
order of magnitude smaller than charm meson production~\cite{cleo:cs}.
Charm baryon-antibaryon pair production occurs at energies above
$4.5\gev$.

Another motivation for studying exclusive open charm production is the
existence of a mysterious family of charmonium-like states with masses
above open-charm threshold and quantum numbers $J^{PC}=1^{--}$.
Although these have been known for over four years, the nature of
these states, found in $\ee\to\ppjpsi(\pp)\gisr$ processes, remains
unclear.  Among them are the \y\ state observed by
BaBar~\cite{babar:y4260,babar:y4260_08}, confirmed by
CLEO~\cite{cleo:y4260_isr,cleo:y4260_scan} and
Belle~\cite{belle:y4260}; the $Y(4350)$ discovered by
BaBar~\cite{babar:y4350} and confirmed by Belle~\cite{belle:y4350};
and two structures, the $Y(4008)$ and the $Y(4660)$ seen by
Belle~\cite{belle:y4260,belle:y4350}.

No clear evidence for open charm production associated with any of
these states has been observed. In fact the \y\ peak position appears
to be close to a local minimum of both the total hadronic cross
section~\cite{bes:cs} and of the exclusive cross section for
$\ee\to\dstdst$~\cite{belle:dst,babar:dd_new}. The $X(4630)$, recently
found in the \eell\ cross section as a near-threshold
enhancement~\cite{belle:x4630}, has a mass and width (assuming the
$X(4630)$ to be a resonance) consistent within errors with those of
the $Y(4660)$, supporting explanation that the $X(4630)$ is
$Y(4660)$~\cite{x4630:bugg} or $\psi(2S)f_0(980)$ bound
state~\cite{x4630:molecule}. However, this coincidence does not
exclude other interpretations of the $X(4630)$, for example, as a
conventional charmonium state~\cite{x4630:charm} or as
baryon-antibaryon threshold effect~\cite{x4630:thresh}, point-like
baryons~\cite{x4630:point}, or as a tetraquark
state~\cite{x4630:tetra}.

The absence of open charm decay channels for $Y$ states, large partial
widths for decay channels to charmonium plus light hadrons and the
lack of available $J^{PC}=1^{--}$ charmonium levels are inconsistent
with the interpretation of the $Y$ states as conventional charmonia.
To explain the observed peaks, some models assign the $3^3D_1(4350)$,
$5^3S_1(4660)$ with shifted masses~\cite{y:ding}, other explore
coupled-channel effects and rescattering of charm
mesons~\cite{voloshin:rescattering}.  More exotic suggestions include
hadro-charmonium~\cite{hadroch}; multiquark states, such as a
$[cq][\overline{cq}]$ tetraquark~\cite{y4260:tetra} and $D \overline
D{}_1$ or \dndst\ molecules~\cite{y4260:molecule}.  One of the most
popular exotic options for the $Y$ states are the hybrids expected by
LQCD in the mass range from $4.2-5.0\gevc$~\cite{y4260:hybryds}.  In
this context, some authors expect the dominant decay channels of the
Y(4260) to be $\y\to D^{(*)}\overline D{}^{(*)}\pi$.

In this paper we report a measurement of the exclusive \eeddstp\ cross
section as a function of center-of-mass energy from the
\ddstp\ threshold to 5.2\gev, as part of our
studies of the exclusive open-charm production in this mass range. The
analysis is based on a data sample collected with the Belle
detector~\cite{det} at the $\Upsilon(4S)$ resonance and nearby
continuum with an integrated luminosity of $695\,\mathrm{fb}^{-1}$ at
the KEKB asymmetric-energy \ee\ collider~\cite{kekb}.

We employ the reconstruction method that was used for \eedd\ and
\eeddp\ exclusive cross section
measurements~\cite{belle:dd,belle:4415}.  We select \eeddstpg\ signal
events by reconstructing the $D^0$, $D^{*-}$ and $\pi^+$ mesons.  In
general the \gisr\ is not required to be detected: instead, its
presence in the event is inferred from a peak at zero in the spectrum
of recoil mass squared against the \ddstp\ system.  The square of the
recoil mass is defined as:
\begin{eqnarray}
\RMS(\ddstp)=(\ecm - E_{\ddstp})^2 - p^2_{\ddstp}
\end{eqnarray}
Here \ecm\ is the initial \ee\ center-of-mass ($\mathrm{c.m.}$)
energy, $E_{\ddstp}$ and $p_{\ddstp}$ are the $\mathrm{c.m.}$ energy
and momentum of the \ddstp\ combination, respectively.  To suppress
backgrounds two cases are considered: (1) the \gisr\ is outside of the
detector acceptance and the polar angle for the \ddstp\ combination in
the c.m. frame is required to be
$|\mathrm{cos}(\theta_{\ddstp})|>0.9$; (2) the fast \gisr\ is within
the detector acceptance ($|\mathrm{cos}(\theta_{\ddstp})|<0.9$), in
which case it is required to be detected and the mass of the
$\ddstp\gisr$ combination must be greater than ($\ecm -0.58\gevc$).
To suppress background from $\ee \to D^0 D^{*-}\pi^+ (2n)\pi^{\pm}
\gisr$ $(n>1)$ processes we exclude events that contain additional
charged tracks that are not used in $D^0$, $D^{*-}$ or $\pi^+$
reconstruction.

All charged tracks are required to originate from the vicinity of the
interaction point (IP); we impose the requirements $dr<1 \,
{\mathrm{cm}}$ and $|dz|<4\,{\mathrm{cm}}$, where $dr$ and $|dz|$ are
the impact parameters perpendicular to and along the beam direction
with respect to the IP.  Charged kaons are required to have a ratio of
particle identification likelihood, $\mathcal{P}_K = \mathcal{L}_K /
(\mathcal{L}_K + \mathcal{L}_\pi)$, larger than 0.6~\cite{nim}.
Charged tracks not identified as kaons are assumed to be pions.

$K^0_S$ candidates are reconstructed from $\pi^+ \pi^-$ pairs with an
invariant mass within $10\mevc$ of the $K^0_S$ mass. The
distance between the two pion tracks at the $K^0_S$ vertex must be
less than $1\,\mathrm{cm}$, the transverse flight distance from the IP
is required to be greater than $0.1\,\mathrm{cm}$, and the angle
between the $K^0_S$ momentum direction and the flight direction in the
$x-y$ plane should be smaller than $0.1\,\mathrm{rad}$.

Photons are reconstructed from showers in the electromagnetic
calorimeter with energies greater than $50 \mev$ that are not
associated with charged tracks. ISR photon candidates are required to
have energies greater than $2.5 \gev$. Pairs of photons are combined
to form $\pi^0$ candidates. If the mass of a $\gamma \gamma$ pair lies
within $15\mevc$ of the  $\pi^0$ mass, the pair is fitted with
a $\pi^0$ mass constraint and considered as a $\pi^0$ candidate.

$D^0$ candidates are reconstructed using five decay modes: $K^-\pi^+$,
$K^-K^+$, $K^-\pi^-\pi^+\pi^+$, $K^0_S\pi^+\pi^-$ and $K^-\pi^+\pi^0$.
A $\pm 15\mevc$ mass window is used for all modes except for
$K^-\pi^-\pi^+\pi^+$ where a $\pm 10\mevc$ requirement is applied
($\sim 2.5\,\sigma$ in each case).  $D^+$ candidates are reconstructed
using $K^- \pi^+ \pi^+$ and $K^0_S\pi^+$ decay modes~\cite{bckg}; a
$\pm 15\mevc$ mass window is used for both $D^+$ modes. To improve the
momentum resolution of $D$ meson candidates, final tracks are fitted
to a common vertex with a mass constraint on the $D^0$ or $D^+$
mass. $D^*$ candidates are selected via the $D^{*+} \to D^0 \pi^+$ and
$D^{*0} \to D^0 \pi^0$ (for background study) decay modes with a $\pm
2\mevc$ $D^*-D$ mass-difference window ($\sim 3\,\sigma$).  A mass-
and vertex-constrained fit is also applied to $D^*$ candidates.

To remove contributions from the $\ee\to D^{*+} D{}^{*-}\gisr$
process, we exclude \dnpip\ combinations with invariant mass within
$\pm 5\mevc$ of the nominal $D^{*+}$ mass.

$D^0$, $D^+$, $D^{*-}$ and $D^{*0}$ mass sidebands are selected for
the background study; these are four times as large as the signal
region and are subdivided into windows of the same width as the
signal. To avoid signal over-subtraction, the selected $D$ sidebands
are shifted by $30\mevc$ ($20\mevc$ for the $D^0\to
K^-\pi^-\pi^+\pi^+$ mode) from the signal region.  The $D$ candidates
from these sidebands are refitted to the central mass value of each
window.  $D^*$ sidebands are shifted by $4\mevc$ to the higher mass
side of the signal region.

\begin{figure}[htb]
\begin{tabular}{cc}
\hspace*{-0.025\textwidth}
\includegraphics[width=0.49\textwidth]{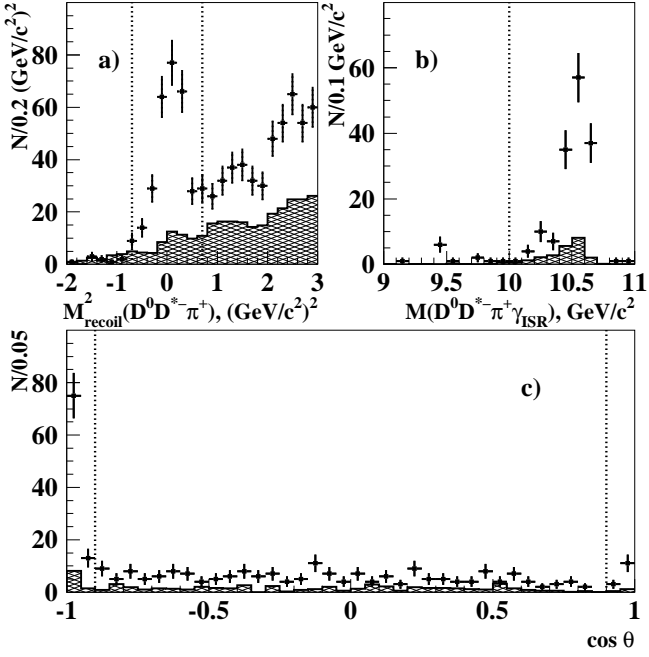}
\end{tabular}
\caption{a) The distribution of $\RMS(\ddstp)$.  b) The mass spectrum
  of $\ddstp\gisr$ combinations. c) The polar angle distribution of
  \ddstp\ combinations.  Histograms show the normalized \mdn\ and
  \mdstm\ sideband contributions. The selected signal windows are
  illustrated by vertical dotted lines.}
\label{fig1}
\end{figure}

\begin{figure}[htb]
\begin{tabular}{cc}
\hspace*{-0.025\textwidth}
\includegraphics[width=0.49\textwidth]{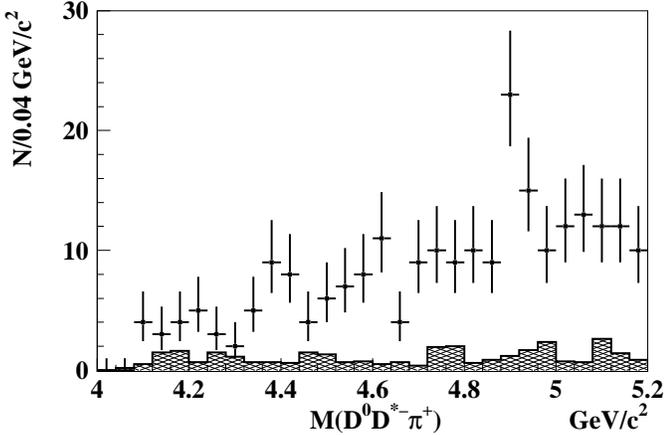}
\end{tabular}
\caption{ The \mddstp\ spectrum.  The histogram shows the
  normalized \mdn\ and \mdstm\ sideband contributions.}
\label{fig2}
\end{figure}

The distribution of $\RMS(\ddstp)$ for the signal region in the data
for $\mddstp<5.2\gevc$ is shown in Fig.~\ref{fig1}\,a). A clear peak
corresponding to the \eeddstpg\ process is evident around zero. The
shoulder at positive values is due to $\ee \to D^0 D^{*-}\pi^+\gisr +
(n)\pi^0 + (m)\gamma $ events.  We define the signal region for
$\RMS(\ddstp)$ by a tight requirement $\pm 0.7(\gevc)^2$ around zero
to suppress the tail from such events.  The invariant-mass
distribution of $\ddstp\gisr$ combinations in the data after the
requirement on $\RMS(\ddstp)$ and the polar angle distribution of
\ddstp\ combinations shown in Fig.~\ref{fig1}\,b), c) are typical of
ISR production and are in agreement with the MC simulation.
The \mddstp\ spectrum obtained after all the requirements is shown in
Fig.~\ref{fig2}.

The contribution of multiple entries after all the requirements is
found to be less than $6\%$.  In such case the single
\ddstp\ combination with the minimum value of $\chi^2_{\mathrm{tot}} =
\chi^2_{M(D^0)} + \chi^2_{M(D^{*-})}$ is chosen, where
$\chi^2_{M(D^0)}$ and $\chi^2_{M(D^{*-})}$ correspond to the mass fits
for $D^0$ and $D^{*-}$ candidates.

The following sources of background are considered:
\begin{itemize}
\item[(1)] combinatorial background under the $D^0$($D^{*-}$) peak
  combined with a real $D^{*-}$($D^0$) coming from the signal or other
  processes;
\item[(2)] both $D^0$ and $D^{*-}$ are combinatorial;
\item[(3)] the reflection from the processes \eeddstppg, where the
  $\pi^0_{\mathrm{miss}}$ is not reconstructed, including $D^{*0}\to
  D^{0}\pi^0_{\mathrm{miss}}$ decays;
\item[(4)] the reflection from the process \eeddstpgg, followed by
  $D^{*0}\to D^0\gamma$, where the low-momentum $\gamma$ is not
  reconstructed;
\item[(5)] \misid\ where the energetic $\pi^0$ is misidentified as a
  single \gisr.
\end{itemize}

The contribution of background (1) is extracted using the $D^0$ and
$D^{*-}$ sidebands.  Background (2) is present in both the \mdn\ and
\mdstm\ sidebands and is, thus, subtracted twice. To take into account
this over-subtraction we use a two-dimensional sideband region, when
events are selected from both the \mdn\ and the \mdstm\ sidebands.
The total contribution from the combinatorial backgrounds (1--2) is
shown in Figs.~\ref{fig1}, \ref{fig2} as a hatched histogram.

Most of the background (3--4) events are suppressed by the tight
requirement on $\RMS(\ddstp)$. The remainder of background (3) is
estimated directly from the data by applying a similar reconstruction
method to the isospin-conjugate process \eeisoch.  Since there is a
charge imbalance in the \dpdstmp\ final state, only events with a
missing extra $\pi^-_{\mathrm{miss}}$ can contribute to the
$\RMS(\dpdstmp)$ signal window.  To extract the level of background
(3), the \dpdstmp\ mass spectrum is rescaled according to the ratio of
$D^-$ and $D^0$ reconstruction efficiencies and an isospin factor of
1/2. A negligibly small contribution of background (3) is found: only
one event with $M_{\dpdstmp}<5.2\gevc$.  Uncertainties in this
estimate are included in the systematic error.  The remainder of
background (4) is estimated from the data assuming isospin
symmetry. We measure the process $\ee\to\dstdstp\gisr$ ($D^{*0}\to
\overline D{}^0\pi^0$) by applying a similar reconstruction
method. Only three events with $M_{\dstdstp}<5.2\gevc$ are found in
the data.  Thus the contribution of background (4) is also found to be
negligibly small; uncertainties in this estimate are included in the
systematic error.

The contribution of background (5) is determined from the data using
fully reconstructed \misid\ events including the reconstruction of an
energetic $\pi^0$. Only one event with $\mddstp<5.2\gevc$ and
$M_{\ddstp\pi^0}>10\gevc$ is found in the data. Assuming a uniform
$\pi^0$ polar angle distribution, this background contribution to the
$|\mathrm{cos}(\theta_{\ddstp})|>0.9$ signal sub-sample (case 1) is
1\,event/9$\eta_{\pi^0} \sim 0.2\,$ events in the entire \mddstp\ mass
range, where $\eta_{\pi^0}$ is the $\pi^0$ reconstruction
efficiency. The probability of $\pi^0 \to \gamma$ misidentification
due to asymmetric $\pi^0 \to \gamma \gamma$ decays is also estimated
to be $\ll 1$. Thus the contribution of background (5) is found to be
negligibly small; uncertainties in this estimate are included in the
systematic error.

\begin{figure}[htb]
\begin{tabular}{cc}
\hspace*{-0.025\textwidth}
\includegraphics[width=0.49\textwidth]{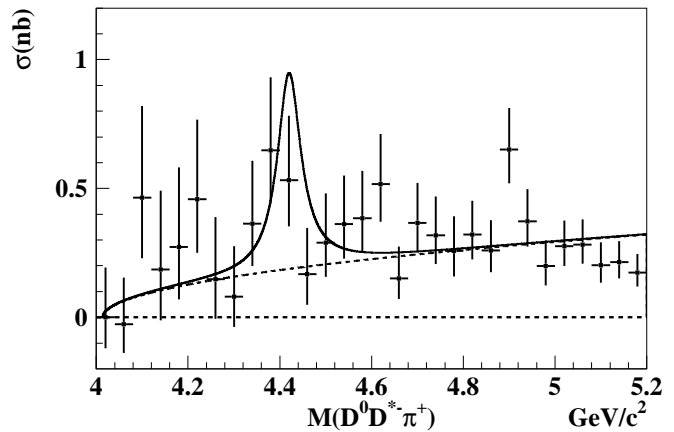}
\end{tabular}
\caption{ The exclusive cross section for \eeddstp\ averaged over the
  bin width with statistical uncertainties only. The fit function
  corresponds to the upper limit on \ps\ taking into account
  systematic uncertainties. The solid line represents the sum of the
  signal and threshold contributions.  The threshold function is shown
  by the dashed line.}
\label{fig3}
\end{figure}

The \eeddstp\ cross section is extracted from the background
subtracted \ddstp\ mass distribution
\begin{eqnarray}
\sigma(\eeddstp) = \frac{ dN/dm }{ \eta_{\mathrm{tot}} dL/dm} \, ,
\end{eqnarray}
where $m\equiv\mddstp$, $dN/dm$ is the mass spectrum obtained without
corrections for resolution and higher-order radiation,
$\eta_{\mathrm{tot}}$ is the total efficiency, and the factor $dL/dm$
is the differential ISR luminosity~\cite{cs}.  The total efficiency
determined by MC simulation grows quadratically with energy from
0.007\% near threshold to 0.036\% at 5.2 \gevc.  The resulting
\eeddstp\ exclusive cross section averaged over the bin width is shown
in Fig.~\ref{fig3} with statistical uncertainties only.  Since the bin
width is much larger than the \mddstp\ resolution, which varies from
$\sim 3\mevc$ around threshold to $\sim 6\mevc$ at $\mddstp=5.2\gevc$,
no correction for resolution is applied.

The systematic errors for the $\sigma(\eeddstp)$ measurement are
summarized in Table ~\ref{tab1}.
\begin{table}[htb]
\caption{Contributions to the systematic error on the \eeddstp\ cross
  section.}
\label{tab1}
\begin{center}
\begin{tabular}{@{\hspace{0.4cm}}l@{\hspace{0.4cm}}||@{\hspace{0.4cm}}c@{\hspace{0.4cm}}}
\hline \hline
Source & Error,[$\%$]
\\ \hline
Background subtraction    & $\pm 3$ \\
Cross section calculation & $\pm 6$ \\
$\mathcal{B}(D)$          & $\pm 3$ \\
Reconstruction            & $\pm 7$ \\
Kaon identification       & $\pm 2$ \\
\hline
Total & $\pm 10$ \\
 \hline \hline
\end{tabular}
\end{center}
\end{table}
The systematic errors associated with the background (1--2)
subtraction are estimated to be 2\% due to the uncertainty in the
scaling factors for the sideband subtractions. This is estimated from
fits to the \mdn\ and \mdstm\ distributions in the data that use
different signal and background parameterizations.  Uncertainties in
backgrounds (3--5) are estimated conservatively to be each smaller
than 1\% of the signal.  The systematic error ascribed to the cross
section calculation includes a 1.5\% error on the differential
luminosity and a 6\% error in the total efficiency function.  Another
source of systematic errors is the uncertainties in track and photon
reconstruction efficiencies (1\% per track, 1.5\% per photon and 5\%
per $K^0_S$). Other contributions come from the uncertainty in the
identification efficiency and the absolute \dn\ and \dstm\ branching
fractions~\cite{pdg}. The total systematic uncertainty is 10\%.

We perform a likelihood fit to the \mddstp\ distribution where we
parameterize a possible \ps\ signal contribution by an $s$-wave
relativistic Breit-Wigner (RBW) function with a free normalization.
We use PDG values~\cite{pdg} to fix its mass and total width.  To take
a non-resonant \ddstp\ contribution into account we use a threshold
function $\sqrt{M-m_{\dn}-m_{\dstm}-m_{\pi^+}}$ with a free
normalization.  Finally, the sum of the signal and non-resonant
functions is multiplied by a mass-dependent second-order polynomial
efficiency function and differential ISR luminosity.

The fit yields $14.4 \pm 6.2 \st_{-9.5}^{+1.0}\sy$ signal events for
the \ps\ state.  The statistical significance for the \ps\ signal is
determined to be $3.1\sigma$ from the quantity $-2 \ln(\mathcal{L}_0 /
\mathcal{L}_{\text{max}})$, where $\mathcal{L}_{\text{max}}$ is the
maximum likelihood returned by the fit, and $\mathcal{L}_0$ is the
likelihood with the amplitude of the RBW function set to zero.  The
goodness of the fit is $\chi^2/n.d.f=1.17$.  The systematic errors of
the fit yield are obtained by varying the mass and total width within
their uncertainties, histogram bin size and the parameterization of
the background function and efficiency.

We calculate the peak cross section for the $\ee\to\ps\to\ddstp$
process at $\ecm = m_{\ps}$ from the amplitude of the RBW function in
the fit to be $\sigma(\ee\to\ps)\times \br( \ps \to \ddstp) < 0.76$ nb
at the 90\% C.L.  Using $\sigma(\ee \to \ps) = 12\pi / m^2_{\ps}
\times (\bee)$ and PDG values of the \ps\ mass, full width and
electron width~\cite{pdg} we found $\bee \times \br(\ps \to \ddstp)<
0.99 \times 10^{-6}$ at the 90\% C.L and $\br(\ps \to \ddstp) <
10.6\%$ at the 90\% C.L. All presented upper limit values include
systematic uncertainties.  For illustration we include the
corresponding fit function on the cross section distribution plot
shown in Fig.~\ref{fig3}.

To obtain limits on the decays $X\to\ddstp$, where $X$ denotes \y,
$Y(4350)$, $Y(4660)$ or $X(4630)$ states, we perform four likelihood
fits to the \mddstp\ spectrum each with one of the $X$ states, the
\ps\ state and a non-resonant contribution.  For fit functions we use
the sum of two $s$-wave relativistic RBW functions with a free
normalization and a threshold function
$\sqrt{M-m_{\dn}-m_{\dstm}-m_{\pi^+}}$ with a free normalization. The
sum of the signal and non-resonant functions is multiplied by the
mass-dependent second-order polynomial efficiency function and
differential ISR luminosity.  For masses and total widths of the
\y\ and \ps\ states we use PDG values~\cite{pdg}. The corresponding
parameters of the $Y(4660)$, $Y(4350)$ and $X(4630)$ states are fixed
from Ref.~\cite{bellebabar:y,belle:x4630}, respectively.

The significances for the \y, $Y(4350)$, $Y(4660)$ and $X(4630)$
signal are found to be $0.9\sigma$, $1.4\sigma$, $0.1 \sigma$ and
$1.8\sigma$, respectively.  The calculated upper limits (at the 90\%
C.L.) on the peak cross sections for $\ee\to X \to \ddstp$ processes
at $\ecm = m_{X}$ are presented in Table~\ref{tab2}. Using fixed
values of $X$ masses and full widths we obtain upper limits on the
$\bee \times \br(X \to \ddstp)$ at the 90\% C.L.  Finally, for the
\y\ state we estimate the upper limit on $\br(\y \to \ddstp)/ \br(\y
\to \ppjpsi)$ at the 90\% C.L. using $\bee \times
\gi(\ppjpsi)$~\cite{pdg}.  For the $Y(4350)$ and $Y(4660)$ states we
calculate $\br( X \to \ddstp)/ \br(X \to\pppsip)$ at the 90\%
C.L. taking into account
$\bee\times\gi(\pppsip)$~\cite{bellebabar:y}. All upper limits
presented in Table~\ref{tab2} are determined by choosing the maximum
signal amplitudes that emerge from: varying masses and widths of the
$X$ states within their uncertainties; varying the histogram bin size;
and changing the parameterizations of the background \& efficiency
functions.
\begin{table*}[htb]
\caption{The upper limits on the peak cross section for the processes
  $\ee\to X\to\ddstp$ at $\ecm = m_{X}$, $\bee \times \br( X \to
  \ddstp)$ and $\br( X \to \ddstp)/ \br(X \to\ppjpsi(\pp)$ at the 90\%
  C.L., where $X = \y$, $Y(4350)$, $Y(4660)$, $X(4630)$.}
\label{tab2}
\begin{center}
\begin{tabular}{l||ccc|c}
\hline
& \y   & $Y(4350)$ & $Y(4660)$ & $X(4630)$ \\\hline
$\sigma(\ee\to X)\times \br(X\to\ddstp)$, [nb] & 0.36 & 0.55 & 0.25 & 0.45 \\
$\bee \times\br( X \to \ddstp)$, [$\times 10^{-6}$] & 0.42 & 0.72 & 0.37 & 0.66 \\\hline
$\br(X \to \ddstp)/ \br(X \to \ppjpsi)$   & 9    &      &      &      \\
$\br(X \to \ddstp)/ \br(X \to \pppsip)$   &      & 8    &   10 &      \\
 \hline \hline
\end{tabular}
\end{center}
\end{table*}

To estimate the effects of possible interference between final states
we also performed a fit to the \mddstp\ spectrum that includes
complete interference between the \ps\ RBW amplitude and a
non-resonant \ddstp\ contribution. We found two solutions both with
$\chi^2/n.d.f=1.28$; the interference is constructive for one solution
and destructive for the other.  From the fit with destructive
interference we find an upper limit on the peak cross section for
$\ee\to\ps\to\ddstp$ process to be $\sigma(\ee\to\ps)\times \br( \ps
\to \ddstp) < 1.93$ nb at the 90\% C.L.

In addition we performed four likelihood fits to the \mddstp\ spectrum
with complete interference between the $X$ and \ps\ states' RBW
amplitudes and a non-resonant \ddstp\ contribution. We found four
solutions for each fit with similar goodness-of-fit
($\chi^2/n.d.f$=1.39, 1.23, 1.39 \& 1.21) and obtained the upper
limits on the peak cross sections for $\ee\to X \to \ddstp$ process to
be $\sigma(\ee\to X)\times \br( X \to \ddstp)$ less than 1.44, 1.92,
1.38 and 0.98 nb at the 90\% C.L. for \y, $Y(4350)$, $Y(4660)$ and
$X(4630)$, respectively.

In summary, we report the first measurement of the \eeddstp\ exclusive
cross section over the center-of-mass energy range from 4.0\gev\ to
5.2\gev. We calculate an upper limit on the peak cross section for the
$\ee\to\ps\to\ddstp$ process at $\ecm = m_{\ps}$ to be 0.76 nb at the
90\% C.L.  The values of the amplitude of the \y, $Y(4350)$, $Y(4660)$
and $X(4630)$ signal function obtained in the fit to the
\mddstp\ spectrum are found to be consistent with zero within errors.
We see no evidence for $\y\to\ddstp$ decays as predicted by hybrid
models and obtain the upper limit $\br(\y \to \ddstp)/ \br(\y \to
\ppjpsi) < 9 $ at the 90\% C.L.

We thank the KEKB group for excellent operation of the accelerator,
the KEK cryogenics group for efficient solenoid operations, and the
KEK computer group and the NII for valuable computing and SINET3
network support.  We acknowledge support from MEXT, JSPS and Nagoya's
TLPRC (Japan); ARC and DIISR (Australia); NSFC (China); DST (India);
MEST, KOSEF, KRF (Korea); MNiSW (Poland); MES and RFAAE (Russia); ARRS
(Slovenia); SNSF (Switzerland); NSC and MOE (Taiwan); and DOE (USA).


\begin{thebibliography} {99}
\bibitem{chcon} Charge-conjugate modes are included throughout this
  paper.
\bibitem{bckg} $D^+$ and $D^{*0}$ candidates are used for the
  background study.
\end{thebibliography}
\end{document}